# Drop of solar wind at the end of the 20th century

**Yuri I. Yermolaev**[1], Irina G. Lodkina[1], Alexander A. Khokhlachev[1], Michael Yu. Yermolaev[1], Maria O. Riazantseva[1], Liudmila S. Rakhmanova[1], Natalia L. Borodkova[1], Olga V. Sapunova[1], and Anastasiia V. Moskaleva[1]

[1]Space Research Institute (IKI RAN), Moscow, Russian Federation (yermol@iki.rssi.ru)

**Abstract**

Variations in the solar wind (SW) parameters with scales of several years are an important characteristic of solar activity and the basis for a long-term space weather forecast. We examine the behavior of interplanetary parameters over 21-24 solar cycles (SCs) on the basis of OMNI database (https://spdf.gsfc.nasa.gov/pub/data/omni). Since changes in parameters can be associated both with changes in the number of different large-scale types of SW, and with variations in the values of these parameters at different phases of the solar cycle and during the transition from one cycle to another, we select the entire study period in accordance with the Catalog of large-scale SW types for 1976-2018 (See the site http://www.iki.rssi.ru/pub/omni, [Yermolaev et al., 2009]), which covers the period from 21 to 24 SCs, and in accordance with the phases of the cycles, and averaging the parameters at selected intervals. In addition to a sharp drop in the number of ICMEs (and associated Sheath types), there is a noticeable drop in the value (by 20-40%) of plasma parameters and magnetic field in different types of solar wind at the end of the 20th century and a continuation of the fall or persistence at a low level in the 23-24 cycles. Such a drop in the solar wind is apparently associated with a decrease in solar activity and manifests itself in a noticeable decrease in space weather factors.

## Introduction

As is well known, the Sun has variable activity, and the study of solar activity variation and its effect on the Earth on a scale of 1-100 years is one of the main tasks of solar and solar-terrestrial physics (e.g., De Jager, 2005; Schwenn , 2006; Hathaway 2015; Schwander et al, 2017; Usoskin, 2017; Christensen-Dalsgaard 2021). In the literature, in addition to the well-studied solar cycles of Schwabe (11 years) and Hale (22 years), the possibility of the Sun approaching the minimum of the Gleissberg (about 90 years) cycle, the so-called grand minimum, is widely discussed (Svalgaard et al., 2005; Zolotova and Ponyavin, 2014; Dreschhoff et al., 2015). These variations in solar activity are manifested along the entire chain of solar-terrestrial connections: in the interplanetary medium, in the magnetosheath, in the magnetosphere, in the ionosphere and in the lower regions of the Earth [Dmitriev et al., 2009; Yermolaev et al., 2012; Li et al., 2016; Miroshnichenko, 2018; Ogurtsov and Jungner, 2020; *Bazilevskaya et al., 2021;* Hajra et al., 2021 and references therein]. In this work, the main attention is paid to the study of the solar wind.

Direct experimental study of the solar wind began at the beginning of the space age and has been one of the main tasks of space physics. There are three main reasons for this attention. First, the large-scale (with scales of more than $10^6$ km at 1 AU) solar wind structures are born on the Sun, do not have time to undergo noticeable modification along the way from the Sun to the Earth and contain information about the structure and processes on the Sun. Secondly, small-scale (with scales less than $10^4$ km) solar wind phenomena are locally induced and allow studying plasma processes in the absence of collisions of charged particles with each other and with the walls of a kind of space laboratory (e.g., Zelenyi & Milovanov, 2004; Verscharen et al, 2019 and references therein). Thirdly, the solar wind is the main agent that transfers disturbances from the Sun to the Earth's magnetosphere and excites perturbations of the magnetosphere-ionosphere system, i.e. solar wind is the causes of various phenomena of space weather (e.g., Tsurutani and Gonzalez 1997; Yermolaev et al., 2005, 2021 and references therein). Thus, the measurement of parameters of the solar wind with different scales is an important method of investigations for the physics of the Sun, the physics of the heliosphere and solar-terrestrial physics.

One of the goals of such studies is to study the variations in solar wind parameters on large time scales exceeding a year. These scales include variations in parameters at different phases of the solar cycle and variations at different cycles. Several similar studies have been carried out (see, for example, Bruno et al, 1994; Dmitriev et al., 2009; Gopalswamy et al. 2015; Li et al., 2016 ; Nakagawa et al., 2019; Larrodera and Cid 2020 and references therein). However, such studies have several disadvantages. (1) A short period of study of the solar wind: usually 2 neighboring cycles are compared, although there are measurements for more than 4 solar cycles, and they are available for analysis (e.g., OMNI base of solar wind parameters https://spdf.gsfc.nasa.gov/pub/data/omni). (2) Lack of selection of the solar wind for its individual types: In most works, if a selection is made according to the types of solar wind streams, then only selection according to the magnitude of the bulk speed (without additional analysis of the connection of these streams with solar structures and/or phenomena), for the so-called fast and slow streams (see, e.g., Bruno et al, 1994; Kasper et al, 2007; Larrodera and Cid 2020 and references therein). This does not allow separating the change in the number of different types of SW and the change in parameters in these different types of SW on these scales.

As shown by numerous experiments at a distance of ~1 AU, the solar wind at scales $> 10^6$ km is structured, i.e. it contains types of streams or regions in which the characteristic parameters either change little or change according to known relations. Some types of streams form on the Sun. These types include both quasi-stationary and disturbed types. Quasi-stationary types are fast (or High Speed Streams (HSS) of solar wind) streams from coronal holes, slow streams from coronal streamers, and also the heliospheric current sheet (HCS), the region of the change in direction of the interplanetary magnetic field (IMF) in the slow type. The disturbed type that is born on the Sun is the body of the coronal mass ejections (CMEs), interplanetary CMEs (ICMEs), which are usually subdivided into magnetic clouds (MC) with high and regularly varying IMF and ejecta with less high and less regular IMF. Another part of the types of streams is formed along the path from the Sun to the Earth. These types are compression regions before the fast stream, corotating interaction region (CIR), which may be formed with the periodicity of the Sun's rotation, since a coronal hole on the Sun can live several revolutions of the Sun, and also the compression region of Sheath, which is formed before the fast ICME. If leading edges of both types of pistons (HSS and ICME) move faster than the preceding solar wind by more than Alfven or sound speed, then shocks form in front of the CIR and Sheath. When the velocity of the trailing edge of the ICME is lower than the velocity of the incident SW stream, a reverse shock wave can be formed in a similar way. If the trailing edge of the ICME or HSS moves faster than the next SW, then a decompression region (the so-called Rarefied region) may form. The Rarefied regions and reverse shocks are formed quite rarely. The origin, formation and dynamics of large-scale SW types are described in detail in the extensive literature (see, e.g., Gosling and Pizzo 1999; Cane and Richardson 2003; Zurbuchen and Richardson 2006;

Jian et al., 2006; Gopalswamy 2006; Yermolaev et al, 2009; 2015; Borovsky and Denton 2016, Kilpua et al., 2017; Regnault et al., 2020 and references therein). Based on the OMNI solar wind measurement database (http://omniweb.gsfc.nasa.gov, King and Papitashvili, 2004 ), we have created a living catalog of large-scale solar wind phenomena since 1976 (the site with web addresses ftp://ftp.iki.rssi.ru/pub/omni/ or http://www.iki.rssi.ru/pub/omni and paper by Yermolaev et al, 2009), and in this work we use the identification of SW types of this catalog.

In this paper, we investigate long-term variations in solar wind parameters on scales from the characteristic sizes of large-scale solar wind structures (several hours or $10^7$ km) to larger sizes, including both variations within the solar activity cycle and variations from the 21st to the 24th solar cycle. In contrast to previous works, for the first time we take into account the change in the number of various large-scale phenomena of the solar wind in different phases and in different cycles of solar activity and perform a separate averaging of the solar wind parameters in the corresponding phenomena.

## 1. Data and Methods

In this work, we use two sources of information.

(1) Hourly data of OMNI base parameters for 1976-2019 (https://spdf.gsfc.nasa.gov/pub/data/omni/low_res_omni , King and Papitashvili, 2004),

(2) Intervals of different types of SW of the catalog of large-scale phenomena (ftp://ftp.iki.rssi.ru/pub/omni/ or http://www.iki.rssi.ru/pub/omni, Yermolaev et al., 2009), created on the basis of the OMNI database.

On the one hand, when creating our catalog, we use the same criteria that were used by other authors to identify types of SW (e.g.Tsurutani etal., 2006; Zurbuchen and Richardson 2006; Wimmer-Schweingruber etal., 2006), and therefore the results are in good agreement with each other (e.g. Lepping et al., 2003; 2017; Jian et al., 2008; Richardson and Cane 2012). On the other hand, unlike other works that identified only individual types of CB, in our catalog each 1-hour measurement point is associated with one of the above types of CB. Due to the small number (less than 2% of the total analysis time) of the rarefaction regions (Borovsky and Denton, 2016), the statistical reliability of the parameters associated with them is low. Therefore, the areas of rarefaction were excluded from the results.

Fig. 1a shows the annual data of the OMNI database that are available for analysis. Before the launch of the solar wind monitors WIND and ACE, in the 21st and 22nd solar cycles (Fig. 1b) the data coverage was about 40%, and in the 23rd and 24th cycles it was about 100%. Figure 1b and Table 1 show the division of the entire 1976-2019 interval into sub-intervals by phases of 21-24 cycles. The solar wind parameters were averaged separately for different types of solar wind and for different phases of solar cycles.

Most of the parameters have a very large scatter, and for them the standard deviation turns out to be close to the average value. However, due to a sufficiently large number of measurements of parameters for various types of the solar wind (with the exception of MC for which the number of points was an order of magnitude less than for other types of SW), the statistical error (i.e., the standard deviation divided by the square root of the number of measurement points) turns out to be an order of magnitude smaller, and the time variations described below on the scales of the phases of solar cycles and cycles are of sufficient statistical significance [Bendat and Piersol, 1971]. The largest scatter is observed for the proton temperature T, and since it has a lognormal distribution [Burlaga and Lazarus 2000, Dmitriev et al., 2009], we averaged the logT value.

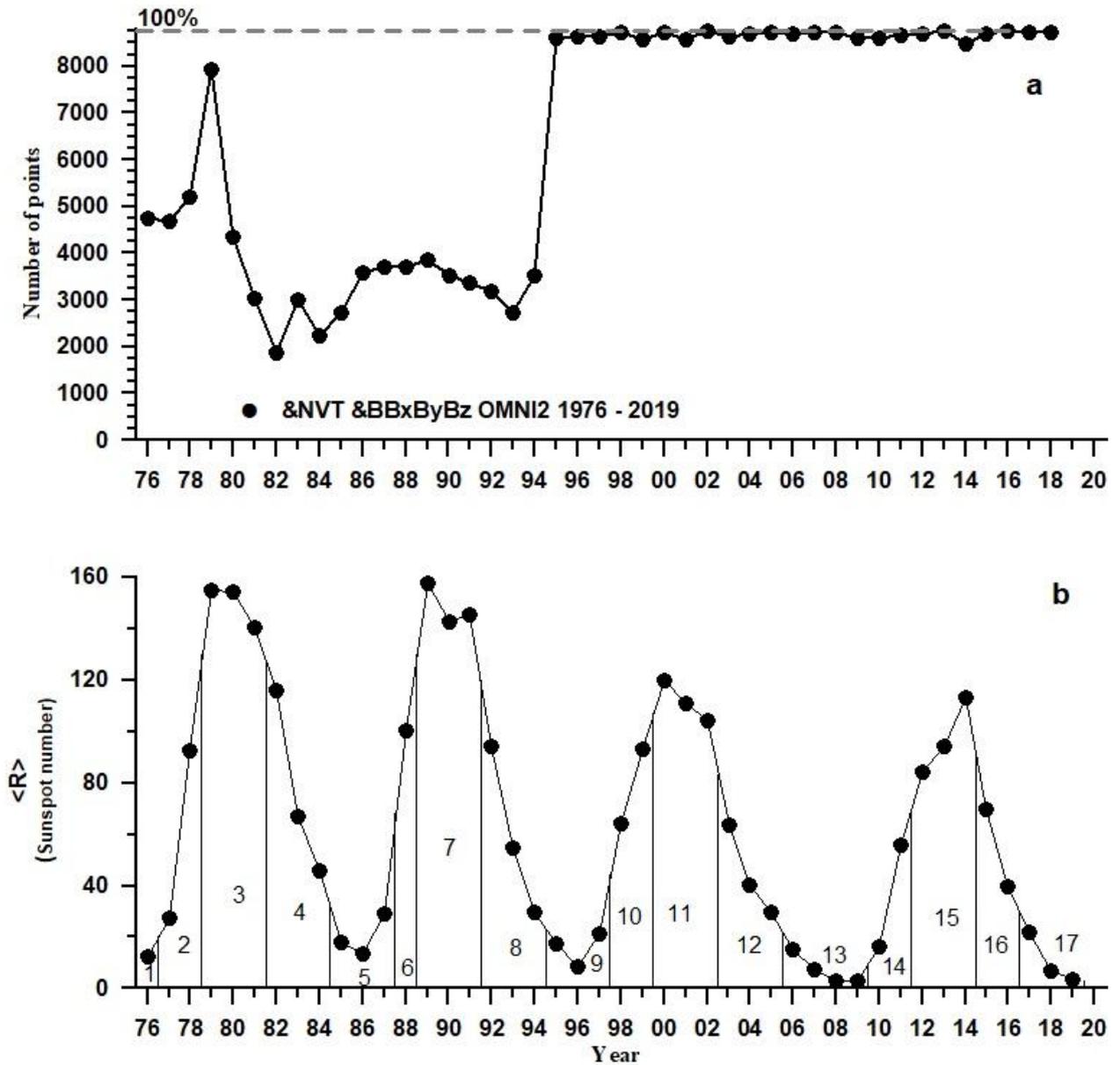

Fig.1. *Annual data a) of the OMNI database, and b) of sunspot number (The numbers and vertical lines show the division over phases of solar cycles 21-24.).*

Table 1. *Averaging intervals over phases of solar cycle*

| № interval | № Cycle | Phase of solar cycle | Years |
|---|---|---|---|
| 1 | 21 | minimum phase | 1976 |
| 2 | | rising phase | 1977,1978 |
| 3 | | maximum phase | 1979-1981 |
| 4 | | declining phase | 1982-1984 |
| 5 | | minimum phase | 1985-1987 |
| 6 | 22 | rising phase | 1988 |
| 7 | | maximum phase | 1989-1991 |
| 8 | | declining phase | 1992-1994 |

| 9  |    | minimum phase  | 1995-1997 |
| 10 | 23 | rising phase   | 1998-1999 |
| 11 |    | maximum phase  | 2000-2002 |
| 12 |    | declining phase| 2003-2005 |
| 13 |    | minimum phase  | 2006-2009 |
| 14 | 24 | rising phase   | 2010,2011 |
| 15 |    | maximum phase  | 2012-2014 |
| 16 |    | declining phase| 2015-2016 |
| 17 |    | minimum phase  | 2017-2019 |

## 3. Results

At the beginning of this section, we present data characterizing solar and magnetospheric activity for the period under study. Then we consider the parameters of the solar wind, selected according to SW types and averaged over the scales of the phases of solar cycles.

### 3.1. A brief overview of solar and magnetospheric activity for the period 21-24 solar cycles

Figure 2 presents an overview of data on solar and magnetosphere activity for the 44-year interval 1976–2019. This interval includes 1635–2225 Carrington rotations of the Sun (~ 27 days) and four maxima of 21–24 solar cycles. The left panel shows the temporal variations in the number of solar X-ray flares of strong M (6344 events) and extreme X (494 events) classes on the visible side of the Sun; the right panel represents moderate (-100 <$Dst_{min}$ <-50 nT; 806 events) and strong ($Dst_{min}$ <-100 nT; 280 events) magnetic storms. The ratio of the number of flares to the number of storms is about seven. After excluding flares located far above 45 ° from the Sun-Earth line, the ratio is approximately three. In the literature, cases are considered when flares of a weaker class C are considered as a possible solar source of disturbances in the magnetosphere. Thus, the number of flares significantly exceeds the number of magnetic storms. This is the main reason for the large number of false alarms for storm forecasts made from observations of solar flares.

    A detailed analysis of solar and magnetospheric data is beyond the scope of this article. Here we would like to draw the reader's attention only to the following facts. These two parameters demonstrate both phase variations within a cycle and a change (a decrease in the number of disturbed events on the Sun and in the Earth's magnetosphere) during a sequential transition from the 21st to the 24th cycle, and when they are averaged over several Carrington rotations, they have a good correlation. However, it is difficult to find a point-to-point correspondence between data on the left and right panels. This confirms the well-known fact that magnetic storms are not directly generated by solar flares (Gosling, 1993 ), and indicates that solar-magnetosphere relationships are more complicated (Gonzalez et al., 1999; Yermolaev et al., 2005).

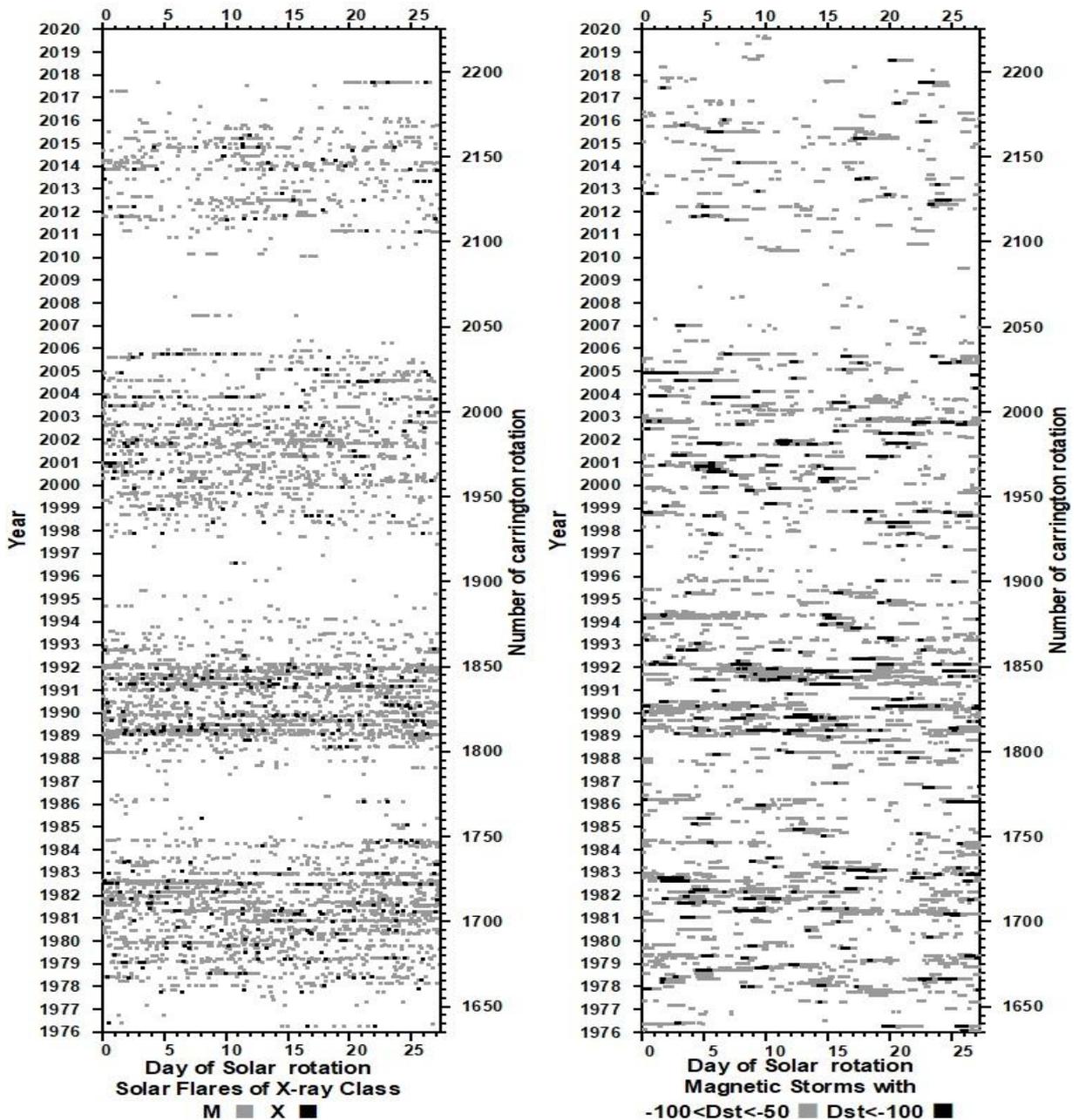

**Fig. 2.** *Time variations in solar X-ray flares (on the left) and magnetic storms (on the right) for the period 1976–2019.*

Figure 3 presents an overview of the disturbed solar wind phenomena for the same time periods of 21-24 solar cycles. The left pane shows the CIR slots (green slots, 1369 events); the right pane shows Sheath (black, 1012 events), Ejecta (blue, 1776 events) and MC (red, 217 events). Gray intervals correspond to other types of solar wind, and white intervals correspond to no measurements. The average annual (averaged over several Carrington rotations) numbers of Sheaths and ICMEs on the right panel demonstrate a similar behavior of the parameters in Fig. 2 with phase variations and a decrease in the number of disturbed events with an increase in the cycle number, while CIRs on the left panel are distributed more evenly in time, without an explicit dependence from the phase of the cycle and without a decrease during 23 and 24 cycles. The relationship between interplanetary drivers and magnetic storms is beyond the scope of this work. Therefore, we note here that the geoeffectiveness (the ratio of the number of magnetic storms to the number of interplanetary drivers of a certain type) for the period 1976-2019 differs little from the geoeffectiveness we obtained earlier for the period 1976-2000 (Yermolaev et al., 2012).

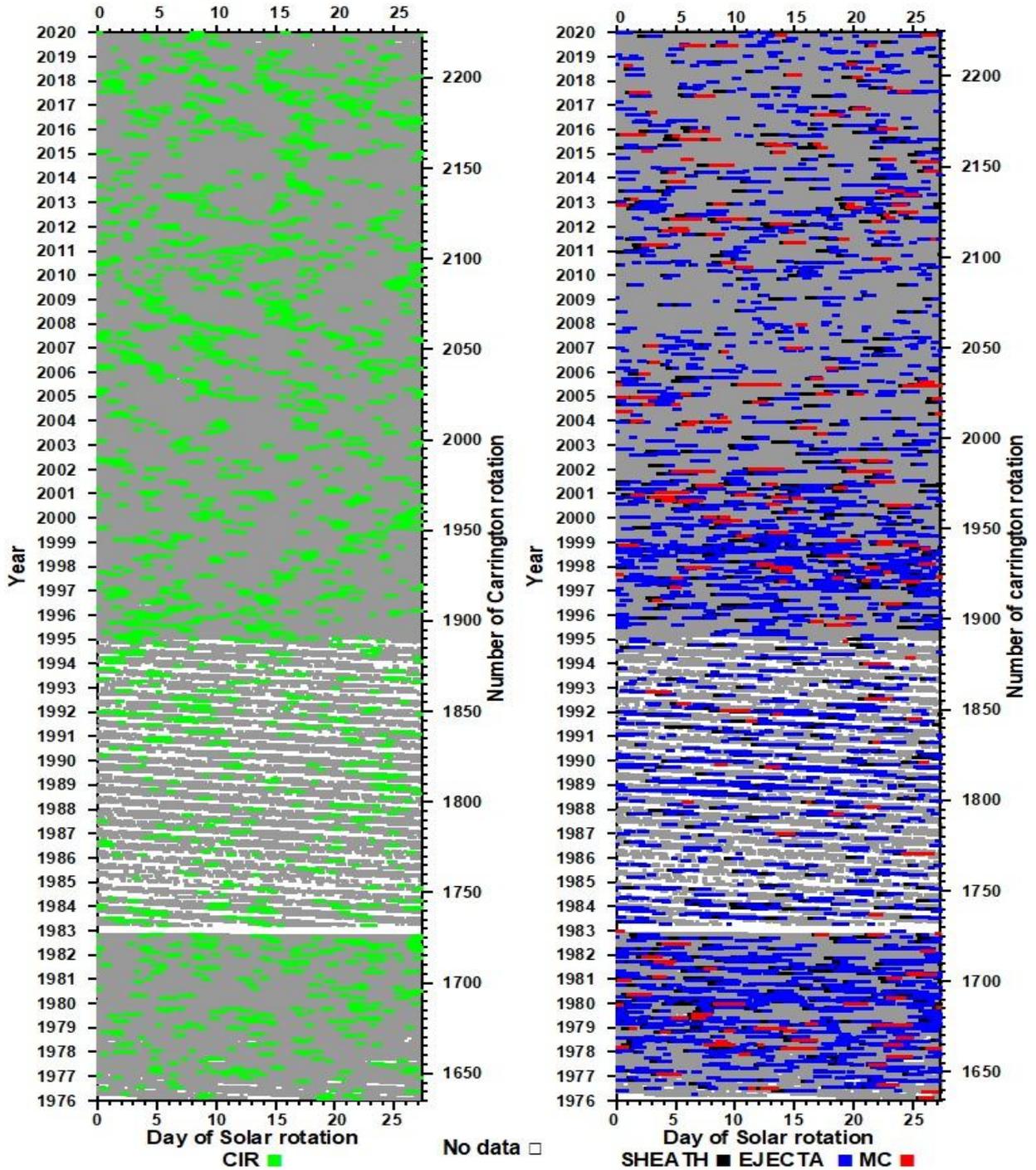

**Fig. 3.** *Time variations in disturbed solar wind phenomena: CIR (on the left) and Sheath, Ejecta, and MC (on the right) for the period 1976–2019.*

### 3.2. Average parameters in solar wind

Table 2 presents several parameters of solar wind and Dst index averaged over solar cycles 21-24 *(top to bottom)*, for three quasi-stationary (HCS, SLOW and FAST) and four disturbed (CIR, Sheath, Ejecta and MC) types of solar wind as well as for ALL (without SW type selection) solar wind data. These cycle-averaged data are shown by red open squares in Figs.4-7 (see below).

*Table 2. Average values and standard deviations of parameters for different types of solar wind in 21-24 solar cycles.*

|  | HCS | SLOW | FAST | ALL (Without SW type Selection) | CIR | Sheath | Ejecta | MC |
|---|---|---|---|---|---|---|---|---|
| N, cm$^{-3}$ | 11,7±6,2<br>14,0±6,5<br>9,1±4,9<br>9,3±5,0 | 9,7±6,5<br>10,1±6,7<br>7,6±5,5<br>7,1±5,1 | 5,4±4,3<br>5,4±3,9<br>4,2±3,5<br>4,0±3,4 | 8,2±6,5<br>8,8±6,7<br>6,8±5,7<br>6,8±5,4 | 11,8±9,0<br>12,6±8,8<br>11,8±7,7<br>11,0±7,4 | 13,1±10,1<br>13,1±8,6<br>11,2±8,3<br>12,4±9,4 | 6,9±4,7<br>6,7±4,7<br>5,2±4,5<br>5,4±3,8 | 9,5±7,1<br>8,5±7,2<br>7,4±6,9<br>6,2±7,0 |
| V*10², km/s | 3,7±0,6<br>3,7±0,5<br>3,7±0,5<br>3,6±0,5 | 3,8±0,4<br>3,8±0,4<br>3,7±0,4<br>3,6±0,4 | 5,4±0,8<br>5,5±0,8<br>5,4±0,8<br>5,3±0,7 | 4,3±0,9<br>4,4±1,0<br>4,4±1,1<br>4,1±0,9 | 4,6±0,9<br>4,6±1,0<br>4,5±0,6<br>4,3±0,8 | 4,7±1,1<br>4,6±1,2<br>4,7±1,2<br>4,3±1,0 | 4,3±0,8<br>4,3±0,9<br>4,5±1,0<br>4,1±0,8 | 4,7±1,1<br>4,7±1,1<br>5,0±1,4<br>4,5±0,9 |
| B, nT | 5,0±2,0<br>4,7±1,9<br>4,4±2,2<br>4,1±1,7 | 6,7±2,9<br>6,7±3,0<br>6,0±2,9<br>5,1±2,6 | 7,5±3,8<br>7,4±3,6<br>6,8±3,9<br>5,6±3,2 | 7,5±3,6<br>7,0±3,6<br>6,3±3,8<br>5,6±3,0 | 9,9±4,1<br>10,2±4,2<br>9,6±2,4<br>8,2±3,1 | 9,9±4,9<br>9,6±4,8<br>10,3±6,1<br>8,3±4,5 | 7,0±2,9<br>7,0±3,0<br>6,6±3,1<br>6,0±2,5 | 13,5±5,1<br>14,3±5,6<br>13,2±7,1<br>11,4±5,9 |
| T/Texp | 1,5±0,7<br>1,3±0,8<br>1,4±0,9<br>1,4±1,0 | 1,6±0,9<br>1,5±1,2<br>1,4±0,8<br>1,3±0,9 | 1,5±0,8<br>1,4±1,0<br>1,3±0,8<br>1,2±0,8 | 1,6±0,9<br>1,4±0,9<br>1,3±0,8<br>1,3±0,8 | 2,3±1,1<br>2,2±1,6<br>2,1±1,0<br>2,1±1,1 | 2,2±1,1<br>2,1±1,3<br>2,0±1,4<br>1,9±1,3 | 1,1±0,8<br>1,0±1,0<br>0,7±0,7<br>0,6±0,4 | 1,1±1,0<br>0,7±0,6<br>0,5±0,7<br>0,5±0,6 |
| LogT, K | 4,77±0,26<br>4,71±0,25<br>4,75±0,29<br>4,70±0,28 | 4,85±0,28<br>4,82±0,29<br>4,78±0,30<br>4,73±0,29 | 5,27±0,26<br>5,27±0,25<br>5,22±0,30<br>5,17±0,27 | 4,91±0,36<br>4,89±0,37<br>4,84±0,39<br>4,76±0,37 | 5,28±0,32<br>5,28±0,35<br>5,23±0,36<br>5,17±0,36 | 5,28±0,33<br>5,24±0,37<br>5,28±0,37<br>5,14±0,39 | 4,87±0,32<br>4,85±0,33<br>4,71±0,34<br>4,57±0,31 | 4,88±0.35<br>4,76±0,32<br>4,63±0,35<br>4,58±0,30 |
| NkT*10$^{-2}$, nPa | 0,91±0,7<br>0,99±0,7<br>0,67±0,5<br>0,60±0,4 | 0,94±0,9<br>0,94±1,2<br>0,64±0,8<br>0,53±0,6 | 1,51±2,0<br>1,59±2,5<br>1,08±2,1<br>0,93±1,7 | 1,19±1,7<br>1,23±1,9<br>0,90±1,8<br>0,74±1,2 | 2,67±2,0<br>2,92±3,1<br>2,24±0,8<br>1,88±1,8 | 3,14±3,6<br>2,90±3,7<br>2,80±5,1<br>2,03±3,7 | 0,61±0,8<br>0,58±1,3<br>0,32±0,5<br>0,26±0,3 | 1,01±1,6<br>0,80±0,9<br>0,47±1,7<br>0,35±0,7 |
| β, 10$^{-1}$ | 13,14±17,0<br>14,88±22,4<br>10,42±14,5<br>11,20±10,3 | 6,69±10,1<br>6,65±12,5<br>5,36±8,2<br>5,95±5,5 | 7,28±6,8<br>7,22±8,6<br>6,09±5,3<br>7,03±4,6 | 6,52±8,8<br>7,14±10,9<br>5,66±6,1<br>6,40±6,7 | 8,07±7,6<br>8,10±10,1<br>6,93±6,4<br>7,53±5,5 | 8,99±8,0<br>8,87±11,4<br>7,43±8,7<br>7,95±7,3 | 4,31±6,5<br>3,89±8,9<br>2,59±6,7<br>2,27±2,5 | 2,08±3,6<br>1,10±2,1<br>1,15±3,9<br>0,89±2,0 |
| Na/Np*10$^{*2}$ | 3,8±2,0<br>3,4±1,6<br>2,9±1,7<br>2,6±1,8 | 4,8±2,8<br>4,5±2,9<br>3,1±1,9<br>2,9±1,9 | 4,9±2,2<br>5,1±2,2<br>4,3±1,9<br>4,1±1,8 | 4,5±2,7<br>4,2±2,8<br>3,3±2,1<br>3,2±2,1 | 5,3±2,4<br>5,3±2,8<br>4,2±2,4<br>3,9±2,3 | 5,0±3,1<br>4,6±3,0<br>3,8±2,9<br>3,5±2,5 | 4,8±3,2<br>4,8±3,2<br>3,4±2,4<br>2,9±2,1 | 6,4±3,7<br>5,9±5,4<br>4,2±3,5<br>5,3±3,6 |
| DsT, nT | -6,7±18,4<br>-3,3±13,9<br>-4,8±15,2<br>-1,2±11,4 | -10,4±19,2<br>-10,3±17,7<br>-8,2±16,3<br>-5,3±14,4 | -27,7±27,8<br>-30,7±25,3<br>-25,6±27,1<br>-18,7±18,7 | -18,9±26,5<br>-18,5±24,9<br>-16,1±25,9<br>-9,3±17,3 | -16,9±27,2<br>-21,1±27,9<br>-12,0±15,6<br>-9,2±21,2 | -21,3±36,2<br>-19,9±32,8<br>-22,8±43,5<br>-8,7±26,4 | -22,0±27,0<br>-23,3±26,6<br>-20,9±24,5<br>-12,1±19,3 | -49,8±50,0<br>-50,8±44,7<br>-62,8±56,2<br>-37,2±39,0 |
| mNV², nPa | 2,7±1,4<br>3,1±1,3<br>1,9±1,0<br>1,9±0,7 | 2,2±1,4<br>2,3±1,5<br>1,7±1,2<br>1,5±1,1 | 2,6±2,4<br>2,7±2,1<br>2,1±2,0<br>1,8±1,6 | 2,4±2,1<br>2,5±1,9<br>2,0±1,9<br>1,8±1,4 | 3,7±2,3<br>4,1±2,5<br>3,5±1,2<br>3,1±1,7 | 4,6±4,3<br>4,1±3,1<br>4,1±3,9<br>3,6±3,1 | 2,0±1,4<br>1,9±1,4<br>1,6±1,3<br>1,4±1,1 | 3,2±2,3<br>3,2±2,7<br>2,9±3,9<br>2,0±2,5 |

Table 2 shows that for all solar cycles the ratios between the parameters averaged over the cycle in different types of SW remain approximately the same. A number of parameters, such as speed V and relative temperature T / Texp, do not change as the cycle number increases. However, all other parameters (concentration N, modulus of IMF B, relative content of helium Na / Np, as well as parameters depending on the IMF, density and temperature - beta parameter, thermal pressure NkT and kinetic pressure mNV², have a clear tendency to decrease. The average amplitude of the DsT index also falls.

Two facts should be noted. On the one hand, in some cases the decline is non-monotonic. This may be due to data gaps in cycle 21 and 22 and the quality of the OMNI database. On the other hand, in most cases, the differences between the mean values exceed the statistical error (except for the cases of MS, which were recorded significantly less than other types of SW).

Figs 4-7 present time profiles of parameters of solar wind plasma and interplanetary magnetic field (IMF) averaged over phases of solar cycles (Table 1 and Fig.1b): minimum – black circles,

rising phase – blue triangles, maximum - purple squares, declining phase - green inverted triangles, without selection with phases – red open squares. Parameters $N, V, B, T/T_{exp}, T$ for 3 quasistationary types of solar wind (HCS, Slow and Fast) and solar wind without type selection and for 4 disturbed types of solar wind CIR, Sheath, Ejecta and MC are shown in Figs. 4 and 6 respectively. Parameters $NkT, \beta, Na/Np, Dst, mNV^2$ are presented in Figs. 5 and 7.

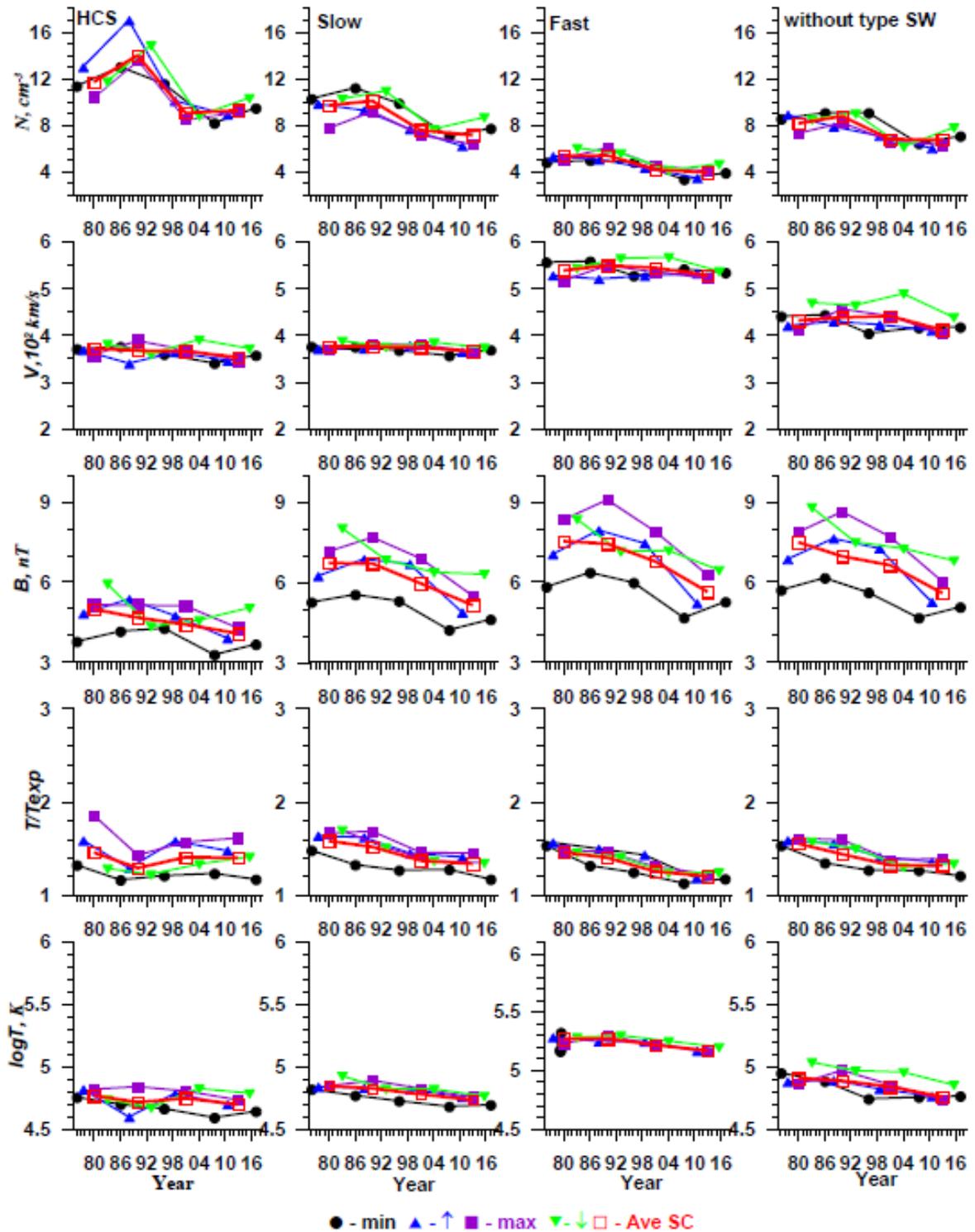

**Fig. 4.** *Time profiles of parameters N, V, B, T / T$_{exp}$, T (top to bottom), averaged over the phases of the solar cycle (see the legend under the figure) for 3 types of solar wind (HCS, Slow and Fast) and solar wind without type selection (see titles at the top of panel columns)*

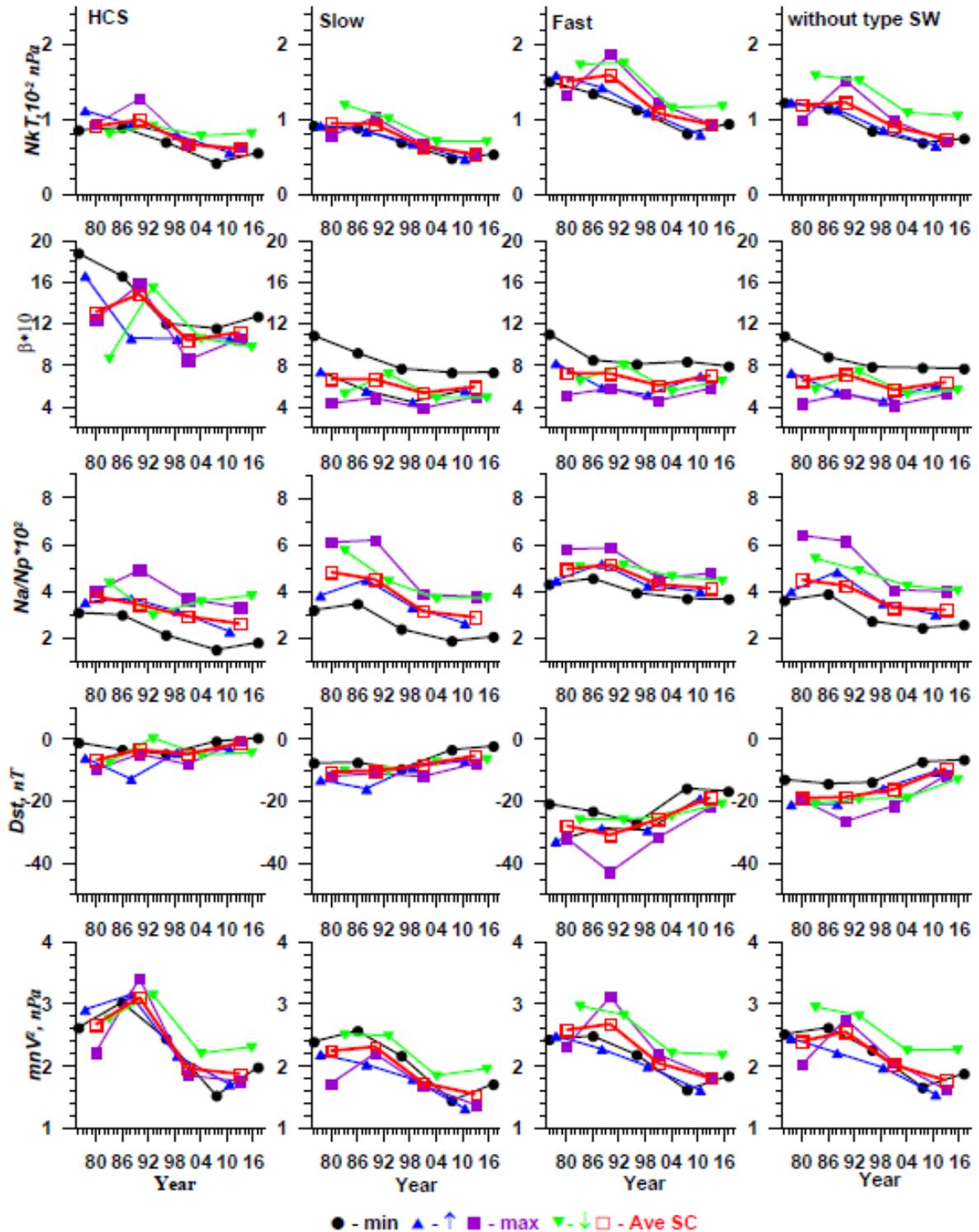

**Fig. 5.** The same as in Fig. 4 for parameters *NkT, β, Na/Np, Dst, mNV²*

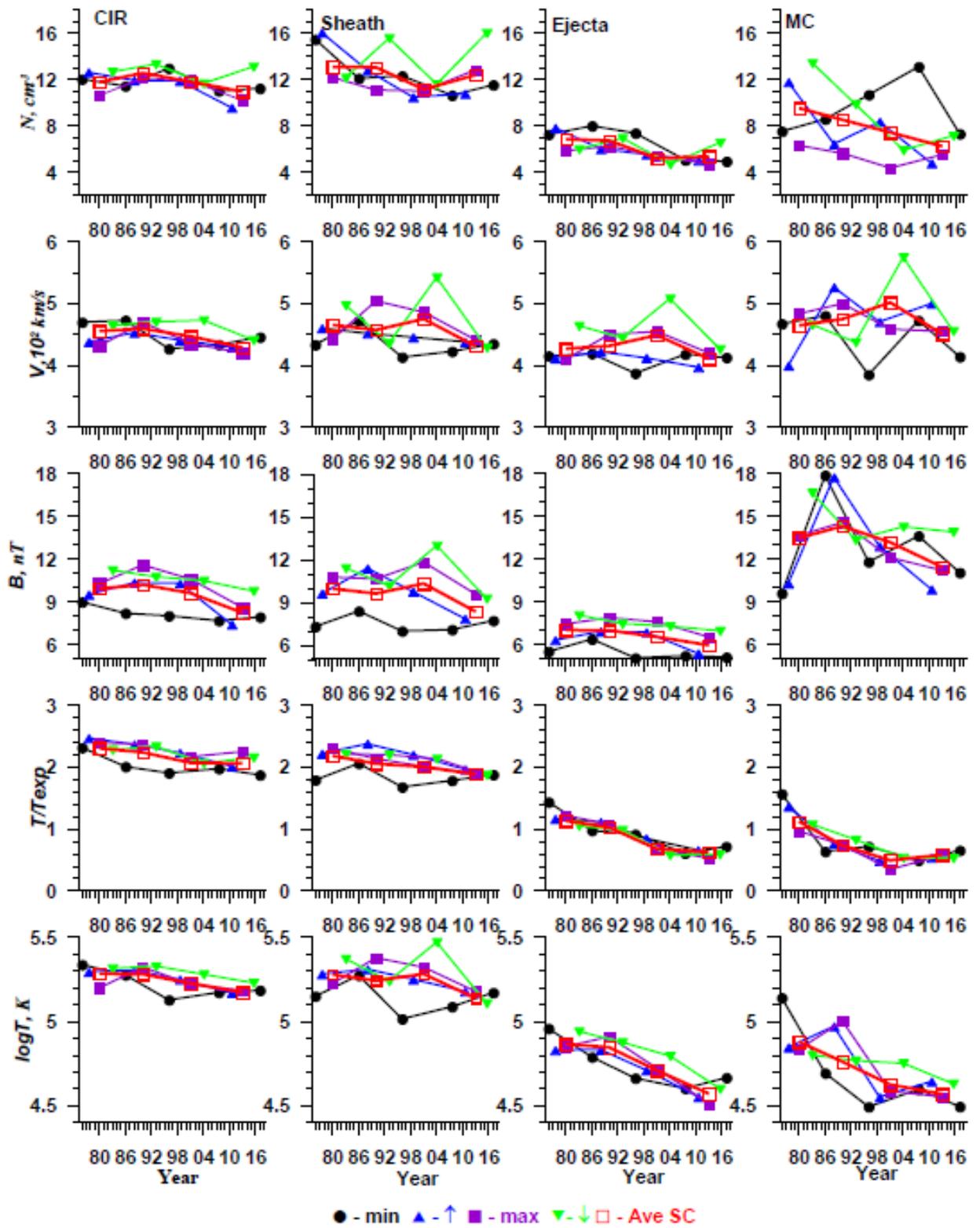

**Fig. 6.** The same as in Fig. 4 for parameters *for 3 types of solar wind* CIR, Sheath, Ejecta and MC

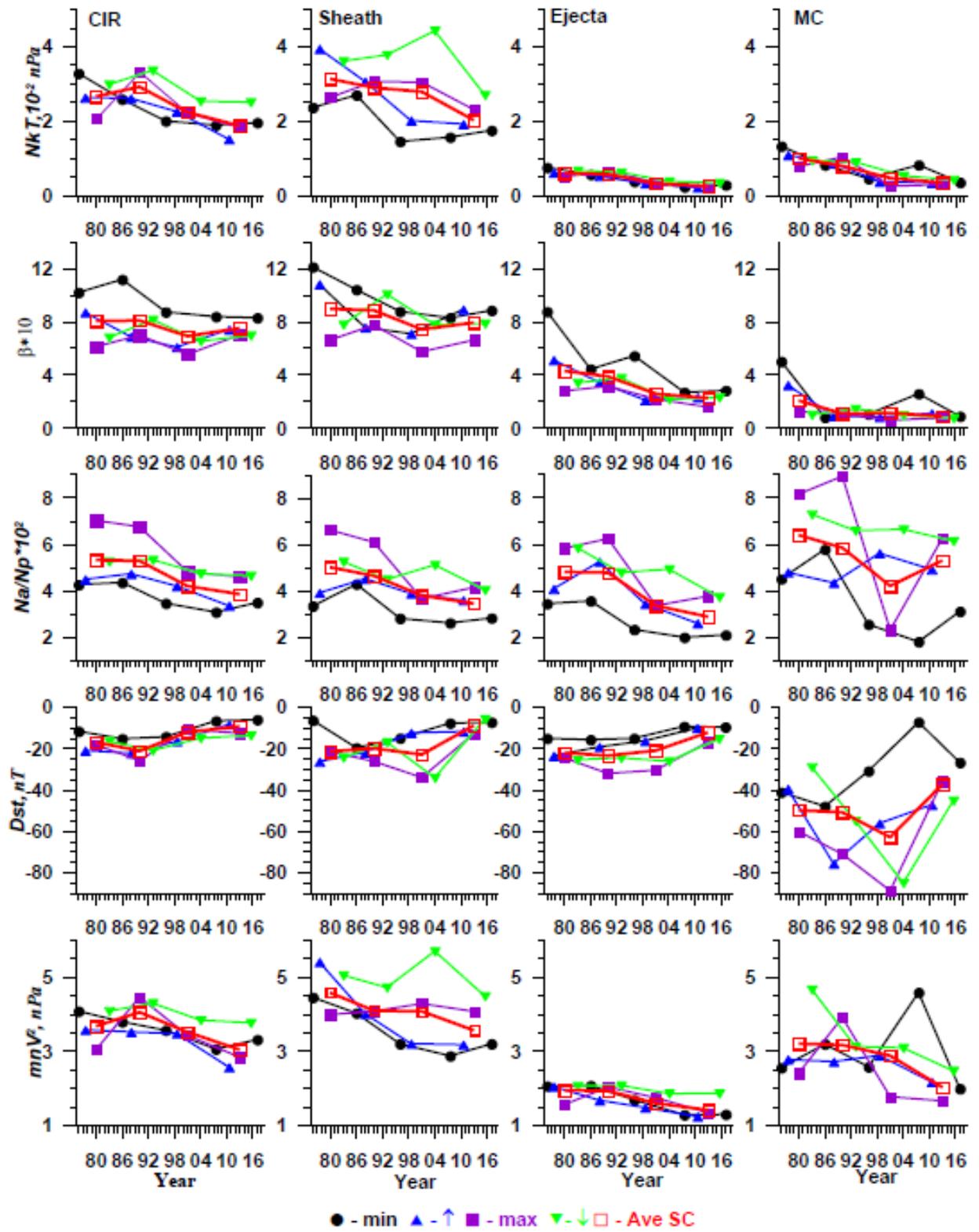

**Fig. 7.** *The same as in Fig. 5 for parameters for 3 types of solar wind CIR, Sheath, Ejecta and MC*

Analysis of the data shows the following.

1) The velocity V in quasi-stationary types of SW changes little both with a change in the cycle phase and with an increase in the cycle numbers. In the disturbed types of SW, the changes are slightly larger. The largest scatter is observed in MC, the number of which is small and the statistical significance is small. However, it can be noted that in Sheath, Ejecta and MC the speed is slightly slower in the minimum phase of cycle 22, and has a maximum in the declining phase of cycle 23.

2) The density N in quasi-stationary types of SW and CIR depends little on the phase of the cycle, while in Sheath and MC there is an increase in the differences for the phases of the cycle: in Sheath, N increased during the declining phase in cycles 22 and 24, while MC has a more chaotic spread , apparently, associated with the small statistics of measurements.

3) The measured and relative temperatures of protons, T and T / TEXP, have little dependence on the phase of the cycle for different types of solar wind, but decreases with an increase in the cycle.

4) The magnitude of the magnetic field B in all types of solar wind depends on the phase of the cycle, has a maximum value at the maximum of the cycle, a minimum at a minimum and an intermediate value in the rising and declining phases.

5) Thermal NkT and dynamic mNV$^2$ pressure, which depend on the parameters V, N and T, behave in a similar way and depend little on the phase of the cycle. Somewhat unexpected was the excess of both pressures in Sheath during the declining phase in all 4 cycles.

6) Parameter β, which is also a dependent parameter on B, N and T, weakly depends on the phase of the cycle, while in all types of SW it is maximum at the minimum and minimum at the maximum of the cycle.

7) The relative helium abundance Na / Np depends on the phase of the cycle: in all cycles, the minimum is observed in the minimum phase, and the maximum value is at the maximum (except for cycle 23, in which the maximum is observed at the declining phase, and a local and rather sharp minimum).

8) The average value of the Dst index weakly depends on the phase of the cycle; nevertheless, there is a slight tendency that the dip of the index is greater (magnetic storms are stronger) at the maximum phase in all types of SW.

## Discussion and Conclusions

Thus, the analysis of the parameters of plasma and IMF was carried out, averaged both by the types of the solar wind and by time on the scales of phases and complete solar cycles from 1976 to 2019 in the plane of the ecliptic near the Earth. Along with the well-known fact that the number of disturbed ICME (and Sheath associated with ICME) solar wind types has significantly decreased over the last 2 cycles, it was shown that CIR events are distributed more evenly in time, without an obvious dependence on the phase of the cycle and without a decrease in for 23 and 24 cycles (see Fig. 3). In addition, it was shown for the first time that most of the parameters for different types of SW decreased noticeably at the end of the 20th century during the transition from the 22nd to the 23rd cycle. In particular, Table 2 and Fig. 4-7 show the time profiles of the solar wind parameters for different types of SW and allow us to draw the following conclusions.

(1) Only the average velocity V did not change over a period of 21-24 cycles in different types of solar wind.

(2) The density N, the absolute T and relative T / Texp temperatures, the magnitude of the IMF B, the relative helium abundance Na / Np, as well as the parameters dependent on the IMF, the density and temperature - beta parameter, thermal pressure NkT and kinetic pressure mNV² decreased markedly (by 20-40% in different types of CB) during the minimum phase between 22 and 23 cycles and remained low in 23-24 cycles. Thus, the activity of the sun and its impact on the heliosphere dropped markedly during this period.

(3) The parameters V, N, T, T / Texp, NkT, $mNV^2$ and β are weakly dependent on the phase of the cycle.

(4) Parameters B and Na / Np differ more strongly in different phases of the cycles and significantly decrease in 23-24 cycles.

(5) A drop in the average IMF B in cycles 23-24 correlates with a decrease in the dip in the Dst index (i.e., with a decrease in the average magnetospheric activity).

Here we discuss in more detail the data on the magnitude of the magnetic field B and the helium content Na / Np in section (4). According to modern concepts (McComas et al., 1992; Webb and Howard, 1994; Svalgaard, and Cliver, 2007; Owens et al., 2008) there is a minimum of the magnetic field, the so-called "The Floor in the Interplanetary Magnetic Field", which is not varies during the solar cycle, and the measured magnetic field is the sum of this minimum field and the ICME field. We estimated the floor in IMF for the 1976-2000 interval, taking into account the contribution of the ICME during the minimum phases of 21 and 22 cycles, as a value of B = 4.65 ± 0.6 nT (Yermolaev et al., 2009b). This value is consistent with earlier estimates: ~ 5 nT (Richardson, Cane, and Cliver, 2002) during 1972-2000, ~ 4.6 nT (Svalgaard and Cliver, 2007) during the last 130 years, but contradicts the estimate 4.0 ± 0.3 nT. obtained at the end of cycle 23 (Owens et al., 2008). Therefore, we similarly made estimates of ~ 4.3 nT for cycle 23 and ~ 4.8 nT for cycle 24 with a standard error for both estimates of ~ 0.7 nT. Thus, our results confirm the decrease in the floor in IMF in cycle 23, obtained in (Owens et al., 2008), but indicate its increase in cycle 24. Due to the uncertainty of the estimates obtained, the question remains whether the fall of the magnetic field B in 23 and 24 solar cycles is associated with a decrease in the floor in the IMF or with a decrease in the number of ICME events and with their contribution to the total magnetic field. Verifying the reliability of these results requires additional research.

An important fact is a significant drop in the helium abundance Na / Np in cycles 23-24 by an amount from ~ 1/4 to ~ 1/3 of the values in the corresponding types of SW in cycles 21-22. Since the change in the chemical composition in the solar wind corresponds to the same change in the composition of the upper atmosphere of the sun, from where the solar wind emanates, this fact can be of great importance for the physics of the sun. In addition, this decrease in the helium abundance may explain its too low content in the MC in papers by Owens, 2018 and Huang et al., 2020, compared with the selection criteria for MC (Na / Np> 0.08), developed on the basis of measurements in cycles 20-22 (Hirshberg et al., 1972; Borrini et al. 1982; Zurbuchen and Richardson [2006]).

It should be noted that although the magnitude of the decrease in parameters during 23 and 24 cycles relative to previous cycles is approximately equal to the magnitude of the change in parameters at different phases of the solar cycle, it exceeds the statistical error and, therefore, is statistically significant. The above results on the decrease in parameters in 23-24 cycles were obtained for different physical quantities measured by various methods and different instruments on various spacecraft included in the OMNI base. This greatly reduces the likelihood that this decline is associated with some methodological effects, and increases the reliability of the results and conclusions drawn.

Thus, it was found that approximately at the minimum phase between the 22nd and 23rd solar cycles, all plasma parameters of the solar wind (with the exception of the velocity) and interplanetary magnetic field fell noticeably in each of the large-scale types of the solar wind and continued to decrease or remain low in 23-24 cycles. A noticeable decrease in magnetospheric activity and other manifestations of space weather is associated with this fact and a decrease in the number and intensity of interplanetary manifestations of CME. Since all measurements were carried out in the plane of the ecliptic, the observed fact can be explained not only by a decrease in the global activity of the Sun, but also by a change in the location of various types of solar wind sources at relatively low and high solar latitudes, which usually occurs when the direction of the solar global magnetic field changes in the solar cycle.

**Acknowledgments**

The authors are grateful to the developers of the OMNI database (http://omniweb.gsfc.nasa.gov). Data on the identification of large-scale types of solar wind for 1976–2020 are available from the site of the Space Research Institute, Moscow, Russia, with web addresses ftp://ftp.iki.rssi.ru/pub/omni/ or http://www.iki.rssi.ru/pub/omni